# The Role of Oxygen in Ionic Liquid Gating on 2D $Cr_2Ge_2Te_6$: a Non-Oxide Material

Yangyang Chen[1,2,†], Wenyu Xing[1,2,†], Xirui Wang[1,2,†], Bowen Shen[1,2], Wei Yuan[1,2], Tang Su[1,2], Yang Ma[1,2], Yunyan Yao[1,2], Jiangnan Zhong[1,2], Yu Yun[1,2], X. C. Xie[1,2], Shuang Jia[1,2]*, and Wei Han[1,2]*

[1]International Center for Quantum Materials, School of Physics, Peking University, Beijing 100871, P. R. China

[2]Collaborative Innovation Center of Quantum Matter, Beijing 100871, P. R. China

†These authors contributed equally to the work

*Correspondence to: weihan@pku.edu.cn (W.H.); gwljiashuang@pku.edu.cn (S. J.)

**ABSTRACT**

Ionic liquid gating can markedly modulate the materials' carrier density so as to induce metallization, superconductivity, and quantum phase transitions. One of the main issues is whether the mechanism of ionic liquid gating is an electrostatic field effect or an electrochemical effect, especially for oxide materials. Recent observation of the suppression of the ionic liquid gate-induced metallization in the presence of oxygen for oxide materials suggests the electrochemical effect. However, in more general scenarios, the role of oxygen in ionic liquid gating effect is still unclear. Here, we perform the ionic liquid gating experiments on a non-oxide material: two-dimensional ferromagnetic $Cr_2Ge_2Te_6$. Our results demonstrate that despite the large increase of the gate leakage current in the presence of oxygen, the oxygen does not affect the ionic liquid gating effect (< 5 % difference), which suggests the electrostatic field effect as the mechanism on non-oxide materials. Moreover, our results show that the ionic liquid gating is more effective on the modulation of the channel resistances compared to the back gating across the 300 nm thick $SiO_2$.

1. INTRODUCTION

Ionic liquid gating has been widely used to induce many interesting physical properties by tuning the carrier density higher than $10^{14}$ cm$^{-2}$ arising from the extremely large electric field on one material.[1-3] For example, superconductivity has been induced by ionic liquid gating in a large variety of materials including $SrTiO_3$, $KTaO_3$, and $La_{2-x}Sr_xCuO_4$, quasi-two dimensional layered ZrNCl, and two-dimensional (2D) transition metal dichalcogenides, etc.[4-12] Besides, the magnetic properties and magnetic phase transition have also been achieved via ionic liquid gating.[13-16] Despite these intensive progresses, whether the gating mechanism is an electrostatic field effect or an electrochemical effect is still under debate, especially for oxide materials, such as $VO_2$ and $SrTiO_3$. For the ionic liquid gating induced metallization in $VO_2$, Nakano et al attributed the mechanism to the carrier doping via electrostatic field effect,[17] while Jeong et al attributed it to the oxygen vacancy formation, which is an electrical chemical process.[18] For $SrTiO_3$,[4, 19-21] Gallagher et al found that the ionic liquid gating can induce the metallization of $SrTiO_3$ protected by a chemically inert hexagonal boron nitride, which supports the electrostatic field effect,[20] while Li et al demonstrated the strong suppression of the metallization in the presence of oxygen, which supports the oxygen vacancy as the major cause.[21]

One of the main experimental proofs to support the electrochemical effect as the mechanism for the metallization in oxide materials is the strong suppression of the ionic liquid gating effect in the presence of oxygen.[18, 21-24] However, in more general scenarios, the role of oxygen in ionic liquid gating effect, i.e. on a non-oxide material, is still not clear. In this letter, we report the negligible role of oxygen in the ionic liquid gating effect on 2D $Cr_2Ge_2Te_6$ (CGT), a non-oxide material. The ionic liquid gating provides a more powerful tool to tune the channel resistance than the back gate through 300 nm thick $SiO_2$. In the presence of oxygen gas, despite the large increase of the gate leakage current, the effect of the ionic liquid gating on the channel resistance is not affected. A tiny difference (< 5 %) of the gating response is observed. Our experimental results indicate that oxygen gas plays a negligible role in the ionic liquid gating. The results are important for future application of ionic liquid and the emerging field of iontronics.

## 2. MATERIALS AND METHODS

**2.1 Materials.** 2D CGT flakes have been shown to be an intrinsically ferromagnetic 2D material recently. [25, 26] To study the ionic liquid gating effect on the 2D CGT device, we synthesized the CGT single crystal by the flux method,[27, 28] and exfoliated the 2D CGT flakes on the $SiO_2$ (300 nm) / Si substrates mechanically using the scotch tape method.[25, 26, 29, 30] The devices are fabricated using standard e-beam lithography and the metallic contacts are made of ~ 80 nm Pt grown by magnetic sputtering.

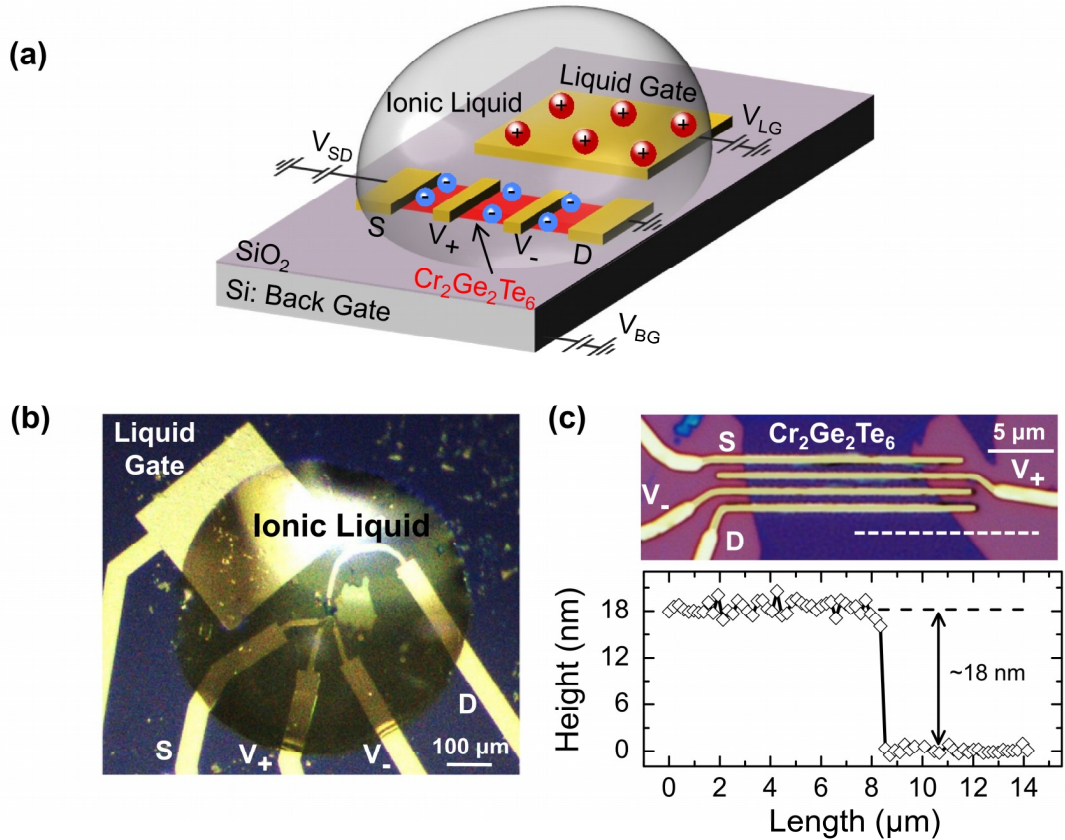

**Figure 1.** Ionic liquid gating on 2D CGT devices. (a) Schematic of the device geometry and the measurement configuration. (b) Optical image of a typical CGT device covered with a droplet of HMIM-TFSI ionic liquid. (c) The magnified image of this CGT device (top) and the CGT thickness of ~ 18 nm determined from the atomic force microscopy (bottom). Four Pt electrodes are contacted to the 2D CGT flake.

**2.2 Device fabrication and measurements.** The details of the synthesis of the CGT single crystals and fabrication of 2D CGT devices can be found in our earlier reports.[26, 28] A lateral gate electrode made of Pt (area: 350 μm × 350 μm) is fabricated near the device as the ionic liquid gate during the same step fabricating the Pt electrical contacts. Prior to the ionic liquid gating experiment, the fabricated devices and ionic liquids are annealed in vacuum (~ 8 × 10$^{-5}$ Pa) at ~ 100 °C for more than 12 hours to remove the water.[18, 21] A ~ 100 nL droplet of ionic liquid, 1-

hexyl-3-methyl-imidazolium bis(trifluoromethylsulfonyl)imide (HMIM-TFSI), is put on the device and the lateral Pt gate electrode, as illustrated in Figures 1 (a) and (b). Then the 2D CGT devices covered with ionic liquid are quickly loaded into a vacuum system (~ $4 \times 10^{-5}$ Pa) for the electrical measurements. Since the CGT devices are fabricated on top of an oxidized Si substrate (300 nm $SiO_2$), the back gate voltage is applied between the highly doped Si and the devices across the 300 nm thick insulating $SiO_2$.

During the gating measurement, a Keithley 2400 source meter is used to apply the liquid gate voltage ($V_{LG}$) and to record the gate leakage current ($I_{LG}$), and another Keithley 2400 source meter is used to apply the back gate voltage ($V_{BG}$) across the 300 nm insulating $SiO_2$. The channel resistance ($R_{XX}$) is obtained from the voltage measured by a Keithley 2002 voltage meter on the center two electrodes ($V_+$ and $V_-$) and the current ($I_{SD}$) measured by another Keithley 2400 source meter that provides a constant voltage ($V_{SD}$) of 0.2 V during the whole measurement.

## 3. RESULTS AND DISCUSSION

**3.1 Ionic liquid gating results.** All the data we present in this letter are taken from the same device (Figure 1 (b) and (c)) to make a fair comparison of the ionic liquid gating response to various oxygen environments. Similar results are obtained on more than 10 CGT devices with various CGT thicknesses. The response of the liquid gating to oxygen gas is the same for all devices that oxygen does not affect the liquid gating on CGT, a non-oxide material. All the measurements on the typical device are performed for at least three times. The thickness of this 2D CGT is ~ 18 nm determined by atomic force microscopy (Figure 1 (c)), and the fabricated 2D CGT device has a channel width of ~ 13.5 μm and a channel length of ~ 1 μm. The ionic liquid gating response of the 2D CGT device in vacuum is measured at first. Figures 2 (a-c) show the time dependence results of $V_{LG}$, $I_{LG}$, and $R_{XX}$ of the 2D CGT device measured at 300 K. The $V_{LG}$ is steadily ramped from 0 V to -0.5 V (orange), -1.0 V (blue), -1.5 V (purple), -2 V (green), -2.5 V (red) with a ramping rate of 10 mV/s, which results in the increase of $I_{LG}$ and decrease of $R_{XX}$. The ionic liquid gate voltage dependence of $R_{XX}$ is showed in Figure 2 (d). The hysteretic behavior of $R_{XX}$ vs. $V_{LG}$ is usually observed in ionic liquid gating studies.[3, 17, 18, 21] Figure 2 (e) summarizes the $R_{XX}$ obtained from the corresponding I-V curves (Inset of Figure 2 (e)) as a function of the ionic liquid gate voltages at 300 K. It is clearly seen that the ionic liquid gating provides a large modulation of channel resistance: $R_{XX}$ at $V_{LG}$ = -2.5 V is ~ 30 times smaller than the value at $V_{LG}$ = 0 V.

**3.2 Ionic liquid gate vs. back gate.** The gating responses from the ionic liquid gate and the back gate across ~ 300 nm thick insulating $SiO_2$ are compared on this device. Figure 3 (a) shows the gating responses for both the back gate voltages ($V_{BG}$) with $V_{LG}$ at 0 V (black stars) and $V_{LG}$ with $V_{BG}$ at 0 V (red balls), and Figure 3 (b) shows the color mapping plot of $R_{XX}$ as a function of $V_{BG}$ and $V_{LG}$. It is clearly showed from Figures 3 (a) and 3 (b) that the liquid gating effect on the CGT is more effective on electric field modulation of the channel resistances compared to the back gating through 300 nm thick $SiO_2$. This observation is consistent with previous reports on the dual gating responses in graphene and $MoS_2$ devices.[3, 8]

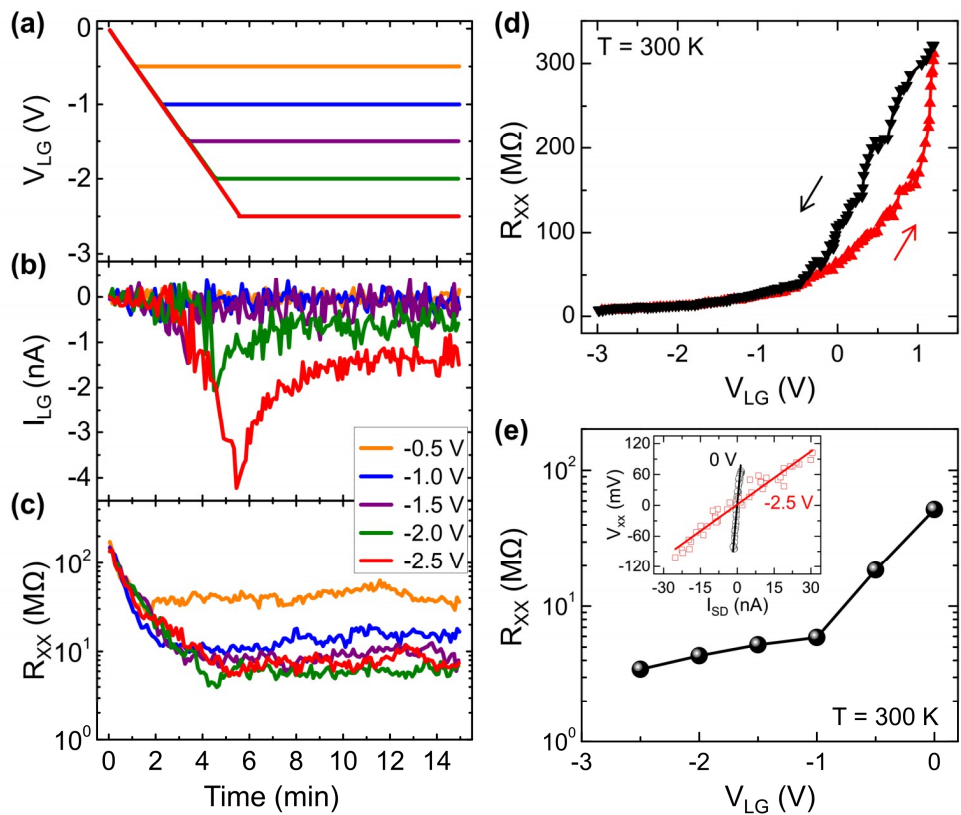

**Figure 2.** Ionic liquid gating response of 2D CGT devices in vacuum. (a-c) Time dependence of $V_{LG}$, $I_{LG}$, and $R_{XX}$ under different gate voltage from -0.5 to -2.5 V. (d) Hysteretic curve of $R_{XX}$ during the ramping measurement of $V_{LG}$ between 1.2 and -3.0 V. (e) $R_{XX}$ vs. $V_{LG}$. $R_{XX}$ is obtained from the current-voltage measurements (inset).

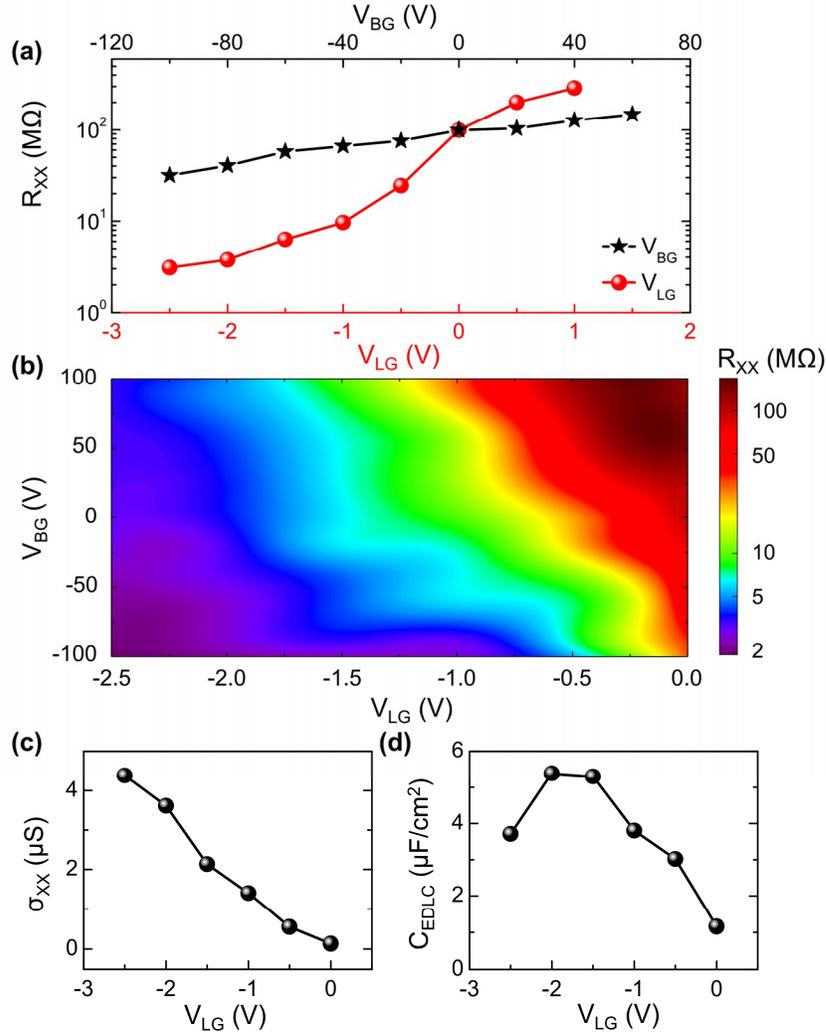

**Figure 3.** Dual gate response of the 2D CGT device. (a) $R_{XX}$ as a function of $V_{BG}$ (black stars) and $V_{LG}$ (red balls). (b) Color mapping of the $R_{XX}$ as a function of the $V_{BG}$ (y axis) and $V_{LG}$ (x axis). (c-d) The channel sheet conductance ($\sigma_{XX}$) and ionic liquid capacitance of ($C_{EDL}$) as a function of $V_{LG}$.

Based on the gate dependence of the channel conductance (Figure 3(c)), the ionic liquid capacitance ($C_{EDL}$) can be calculated using the following equation:[3, 31]

$$C_{EDL} = \frac{d\sigma/dV_{LG}}{d\sigma/dV_{BG}} \frac{\mu_{BG}}{\mu_{LG}} C_{BG} \qquad (1)$$

where $\mu_{BG}$ and $\mu_{LG}$ are the electron field mobility, and $C_{BG}$ is the capacitance for the 300 nm SiO$_2$. Using $C_{BG}$ = 11.5 nF/cm$^2$ from previous studies, the values of $C_{EDL}$ are in the range between 1.2 and 5.4 μF/cm$^2$ as shown in the Figure 3 (d). The obtained values of $C_{EDL}$ are in good agreement with the effective gate capacitances reported in previous ionic liquid gating measurements (1 ~ 20 μF/cm$^2$) on graphene and MoS$_2$.[3, 31] Moreover, based on the gate-tuned channel resistance, the estimated electron field mobility is ~ 0.4 cm$^{-2}$/V·s, which is much lower compared to semiconducting 2D materials, including black phosphorus and MoS$_2$.[32-34] This low mobility might be related to defects scattering in the CGT flakes, which needs further experimental and theoretical studies.

**3.3 Role of oxygen in Ionic liquid gating.** Next, we focus on the study of role of oxygen in ionic liquid gating by performing the gating experiments while systematically varying the oxygen pressure via a high precision leak valve and a turbo pump. Prior to leaking high purity oxygen gas (>= 99.999%) into the vacuum system, we thoroughly pump out and purge the gas line with the oxygen gas to remove any water or impurity gas. Figure 4 (a) shows the ionic liquid gate voltage dependence of R$_{XX}$ in vacuum (~ 4 × 10$^{-5}$ Pa), 10 Pa, and 50 Pa oxygen environments, and Figure 4 (b) shows the corresponding gate leakage current as a function of ionic liquid gate voltage. Obviously, the leakage current is much larger in the presence of oxygen, which is consistent with previous reports on the ionic liquid gating experiments on VO$_2$ and SrTiO$_3$ in oxygen gas.[18, 21] However, the ionic liquid gate voltage dependences of R$_{XX}$ are almost identical under different oxygen partial pressures. The difference of the gating effect is less than 5 % between in vacuum and 50 Pa O$_2$. These results show that the ionic liquid gating of 2D CGT does not affected by the oxygen gas, despite the presence of oxygen largely increases the gate leakage current.

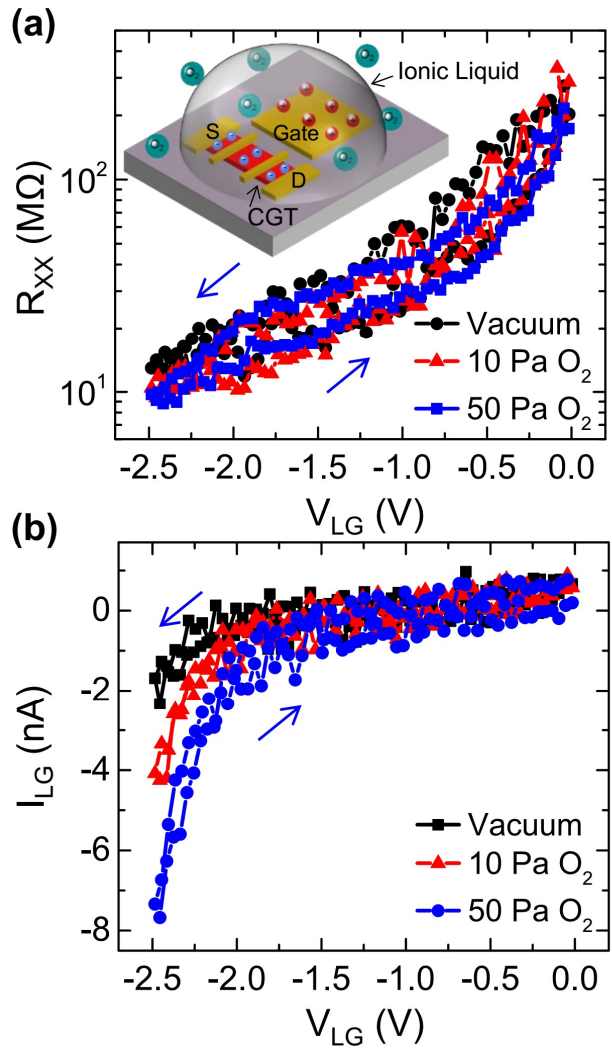

**Figure 4.** Ionic liquid gate measurements of the 2D CGT device in vacuum and oxygen. Hysteretic loops of $R_{XX}$ (a) and $I_{LG}$ (b) as a function of $V_{LG}$ in vacuum ($4 \times 10^{-5}$ Pa, black) and different oxygen pressure (10 Pa, red and 50 Pa, blue). Inset of (a): Schematic of ionic liquid gating on 2D CGT in the presence of oxygen gas.

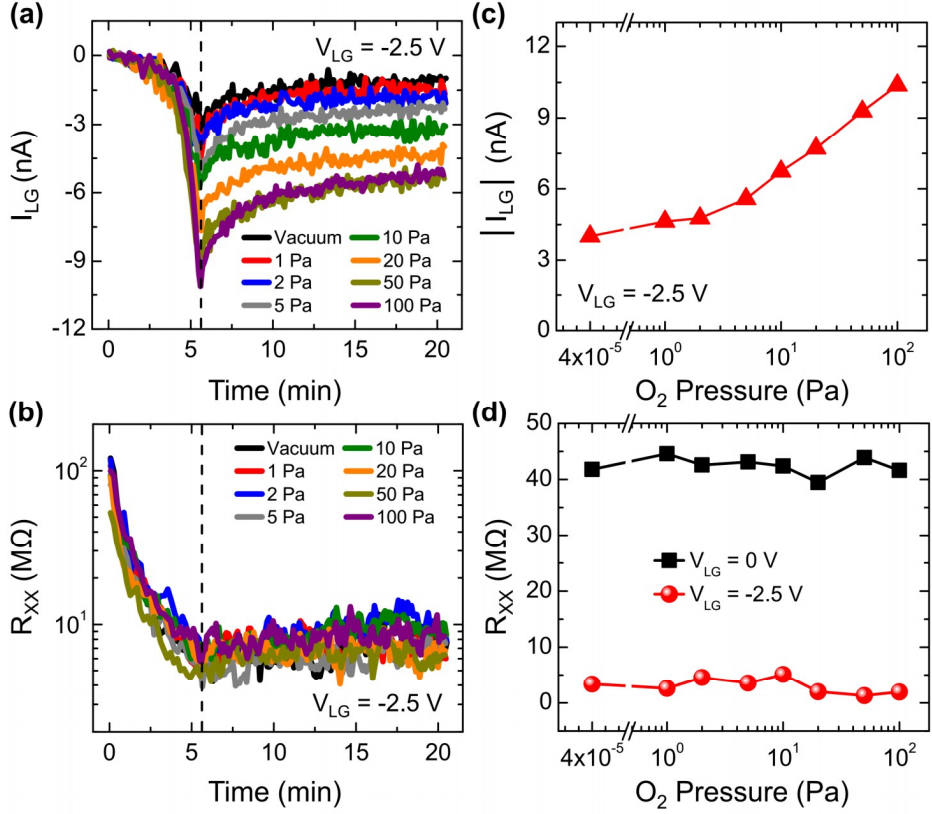

**Figure 5.** The role of oxygen in the liquid gating on the 2D CGT device. (a-b) Time dependence of $I_{LG}$ (a) and $R_{XX}$ (b) at $V_{LG}$ = -2.5 V in vacuum ($4 \times 10^{-5}$ Pa) and different oxygen environments (oxygen pressure from 1 to 100 Pa). (c) The maximum of leakage current vs. oxygen pressure with $V_{LG}$ at -2.5 V. (d) $R_{XX}$ vs. oxygen pressure with $V_{LG}$ at 0 V and -2.5 V.

Then, we systematically change the oxygen pressure and study its effect on the gate leakage current and $R_{XX}$ as a function of time under the same ionic liquid gate voltage of -2.5 V. Figure 5 (a) and 5 (b) show the detailed time dependence of gate leakage current and $R_{XX}$ in various oxygen gas environments (vacuum and 1, 2, 5, 10, 20, 50, and 100 Pa oxygen gas). A larger leakage current corresponds to a higher oxygen pressure, which is consistent with previous reports.[18, 21] The maximum leakage current as a function of the oxygen pressure for $V_{LG}$ = -2.5 V is summarized in Figure 5 (c). The gate leakage current monotonically increases as the oxygen pressure increases. Figure 5 (d) shows the summary of $R_{XX}$ at $V_{LG} = 0$ V and $V_{LG} = -2.5$ V as a function of the oxygen pressure. Despite the large enhancement of the gate leakage current, the gating response of the channel resistances is not affected. These experimental results further support previous studies that the ionic liquid gate induced metallization in various oxide materials ($TiO_2$, $VO_2$, $WO_3$, $SrTiO_3$, etc) arises from the oxygen vacancies.[18, 21-23, 35]

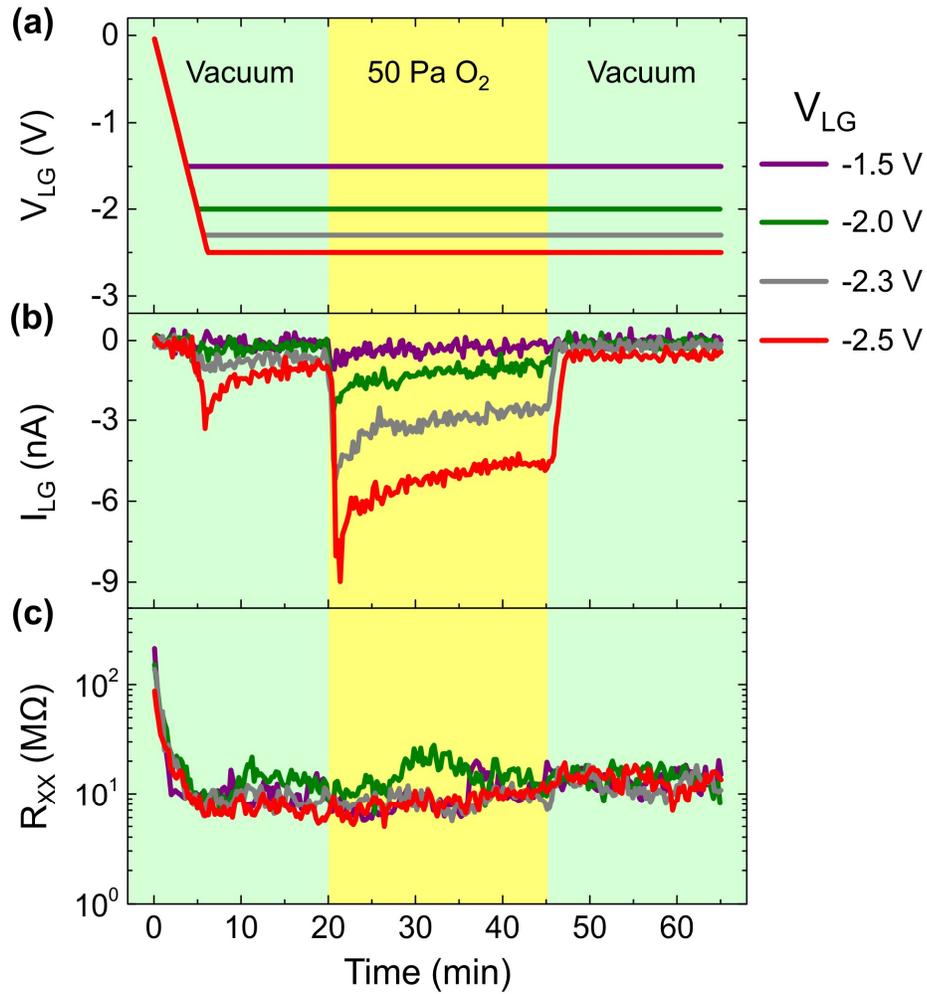

**Figure 6.** The role of oxygen in the liquid gating on the 2D CGT device. Time dependence of $V_{LG}$ (a), $I_{LG}$ (b), and $R_{XX}$ (c) in the cycle of vacuum, oxygen injection, and pump-out back to vacuum for $V_{LG} = -1.5, -2.0, -2.3$, and $-2.5$ V, respectively.

To further identify the negligible role of oxygen in the ionic liquid gating on the 2D CGT device, we measure the response of the leakage current and $R_{XX}$ as the oxygen pressure varies during the gating experiments under the constant ionic liquid gate voltages of -1.5, -2.0, -2.3, and -2.5 V, respectively. As shown in Figure 6, for each ionic liquid gate voltage, 50 Pa oxygen is injected to the vacuum chamber for ~ 25 minutes and then pumped out. As soon as the 50 Pa oxygen is introduced, the leakage current abruptly increases and then decays slowly. When the oxygen gas is pumped out at ~ 45 minutes and the pressure quickly goes back to ~ $4 \times 10^{-5}$ Pa, the gate leakage current dramatically decreases to the value prior to the injection of oxygen gas. The average resistances at $V_{LG}$ = -2.5 V are ~ 9.0 MΩ before oxygen is injected (from 5 to 20 minutes), ~ 8.7 MΩ when the oxygen is maintained in the chamber (from 20 to 45 minutes), and ~ 10 MΩ when the oxygen is pumped out (from 45 to 60 minutes). Compared to the initial resistance at $V_{LG}$ = 0 V (~ 140 MΩ), the variation of the gating response is negligible (< 5%). These results further prove that oxygen does not affect the ionic liquid gating on the 2D CGT device, but only affect the gate leakage current.

## 4. CONCLUSIONS

In summary, we have performed the ionic liquid gating experiments on the 2D ferromagnetic CGT flakes, and have observed a much larger gating effect compared to the back gating across 300 nm thick $SiO_2$. Most importantly, we have demonstrated that role of oxygen in the ionic liquid gating on 2D CGT is negligible despite the large enhancement of gate leakage current in the presence of oxygen gas up to 100 Pa. These results support that the ionic liquid gating effect on $Cr_2Ge_2Te_6$, a non-oxide material, is an electrostatic effect. Our work demonstrates that ionic liquid gating can also be used in oxygen environment for non-oxide systems, which will be helpful for future applications of ionic liquid gating in the field of iontronics.[1]

## AUTHOR INFORMATION

**Corresponding Authors**

*Email: weihan@pku.edu.cn (W.H.); gwljiashuang@pku.edu.cn (S. J.)

Author Contributions

The manuscript was written through contributions of all authors. All authors have given approval to the final version of the manuscript.

[†]These authors contributed equally.

**Funding Sources**

National Basic Research Programs of China and National Natural Science Foundation of China

**Competing interests**

The authors declare no competing financial interests.

## ACKNOWLEDGMENTS

We acknowledge the financial support from National Basic Research Programs of China (973 program Grant Nos. 2015CB921104, 2014CB920902, 2013CB921901 and 2014CB239302) and National Natural Science Foundation of China (NSFC Grant No. 11574006). W.H. also acknowledges the support by the 1000 Talents Program for Young Scientists of China.

**TOC Graphic**

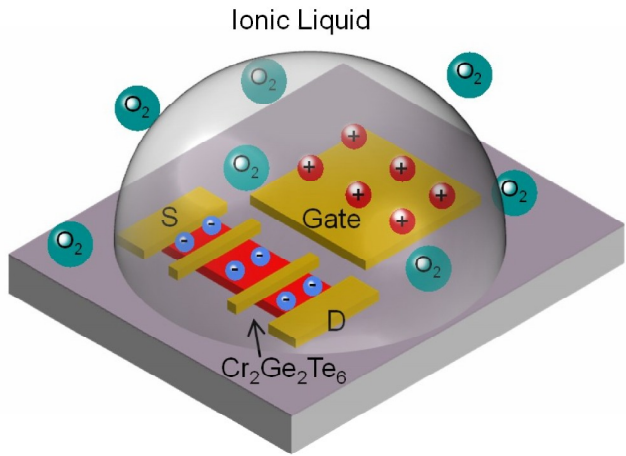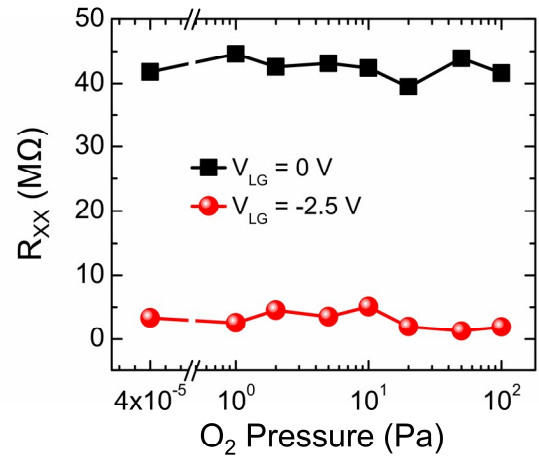

18